\documentstyle[12pt,preprint,aps,prb,eqsecnum]{revtex}
\def\nm{\nonumber}
\def\beqa{\begin{eqnarray}}
\def\beq{\begin{equation}}
\def\F{{\cal{F}}}

\def\eeqa{\end{eqnarray}}
\def\eeq{\end{equation}}
\def\lab{\label}    
\def\pa{\partial}

\def\Del{\Delta}

\begin{document}

\begin{titlepage}
\thispagestyle{plain}
\pagenumbering{arabic}
%%%%%%%%%%%%%%%%%%%%%%%%%%%%%%%%%%%%%%%%%%%%%%%%%%%%%
\begin{center}
{\Large \bf Instanton Correction of Prepotential in}
\end{center}
\vspace{-8.0mm}
\begin{center} 
{\Large \bf Ruijsenaars Model Associated with}
\end{center} 
\vspace{-8.0mm}
\begin{center}
{\Large \bf $N=2$ SU(2) Seiberg-Witten Theory}
\end{center} 
%%%%%%%%%%%%%%%%%%%%%%%%%%%%%%%%%%%%%%%%%%%%%%%%%%%%%%%%%
\lineskip .80em
\vskip 4em
\normalsize
\begin{center}
{\large Y\H uji Ohta}
\end{center}
\vskip 1.5em
\begin{center}
{\em Research Institute for Mathematical Sciences }
\end{center}
\vspace{-11.0mm}
\begin{center} 
{\em Kyoto University}
\end{center}
\vspace{-11.0mm}
\begin{center}
{\em Sakyoku, Kyoto 606, Japan.}
\end{center}
%%%%%%%%%%%%%%%%%%%%%%%%%%%%%%%%%%%%%%%%%%%%%%%%%%%%%%%%%%
\begin{abstract}

Instanton correction of prepotential of one-dimensional SL(2) Ruijsenaars 
model is presented with the help of Picard-Fuchs equation of 
Pakuliak-Perelomov type. It is shown that the instanton induced prepotential 
reduces to that of the SU(2) gauge theory coupled with a massive adjoint 
hypermultiplet. 
\end{abstract}
%%%%%%%%%%%%%%%%%%%%%%%%%%%%%%%%%%%%%%%%%%%%%%%%%%%%%%%%%%%%%%%%%
\end{titlepage}

\pagestyle{myheadings}
\markboth{Prepotential in Ruijsenaars model}
{Prepotential in Ruijsenaars model}

%%%%%%%%%%%%%%%%%%%%%%%%%%%%%%%%%%%%%%%%%%%%%%%%%%%%%%%%%%%%%%%%%%%%%
\begin{center}
\section{Introduction}
\end{center}

The low energy effective action of $N=2$ supersymmetric Yang-Mills theory 
is described by a prepotential and if this is 
obtained we can know (all) informations concerning to the effective theory, 
namely, in this case the theory becomes ``solvable''. Actually, instantons 
have been known to contribute to the prepotential, \cite{Sei} 
but not so much discussions were made. However, 
Seiberg and Witten \cite{SW} pointed out several years ago that it 
was possible to determine the prepotential including instanton effects with 
the help of a Riemann surface and periods of a meromorphic 1-form on it. 
This approach using a Riemann surface is in general referred to 
Seiberg-Witten theory. 

The effective action of gauge theory occasionally including massive 
hypermultiplets in the fundamental representation of the gauge group has 
been discussed in many view points, and accordingly, we have now much 
acquaintance with various properties of the prepotential and it's related 
materials, such as Picard-Fuchs equations for periods, \cite{IMNS,Ali,IS,O3,O4} 
renormalization group like equation for 
prepotential, \cite{Mat,STY,EY,HW,KO,DKP} relation to integrable 
systems, \cite{GKMMM,NT} appearance of WDVV equations, \cite{MMM,MMM2,IY2} 
flat coordinates \cite{IY,IXY} and so on.
 
On the contrary, for a theory coupled with 
adjoint hypermultiplets, \cite{DW} 
not so much compared to the above cases including fundamental 
hypermultiplets are revealed, but there are strong supports that the 
relevant Seiberg-Witten solutions must be related with the integrable 
Calogero dynamical systems, \cite{IM} and in fact, a few examples of 
prepotential associated with spectral curves expected from Calogero 
systems showed a good prediction of the instanton correction. \cite{DP} 

On the other hand, recently, Braden {\em et al.} \cite{BMMM} analyzed a more 
general integrable system, called Ruijsenaars model, which can be thought as a 
``relativistic'' version of Calogero system. \cite{Rui} The Ruijsenaars model 
itself is considered as a candidate of integrable system related to five-dimensional 
gauge theory, \cite{Nek} and it's specific limits are known to recover Seiberg-Witten 
solutions in four and five dimensions. At the perturbative level, we can easily 
establish these correspondences, but we can not conclude that the Ruijsenaars model 
is in fact the integrable system relevant to these gauge theories unless 
instanton contribution is correctly taken into account. 
Therefore, the only necessary item to be discussed 
is now an exact treatment of this model including instanton 
effects. From this reason, we derive the instanton contribution to the 
prepotential of Ruijsenaars model in this paper. 

The paper is organized as follows. In Sec.II, we briefly review the 
correspondence among the Seiberg-Witten solution to the SU(2) gauge theory 
and Calogero and Ruijsenaars systems. In Sec.III, we consider the 
Picard-Fuchs equations for these integrable systems, but actually they are 
found to be available from Pakuliak-Perelomov equations. \cite{PP} In Sec. 
IV, the period integrals and the effective coupling constant of the 
Ruijsenaars model are evaluated in the weak coupling regime. 
In Sec. V, a differential equation for the prepotential is constructed from 
the Pakuliak-Perelomov type Picard-Fuchs equation. One-instanton correction 
of the prepotential is presented. As a check, we consider a 
reduction to the prepotential of Calogero system 
and show that the instanton induced prepotential is consistently determined. 
Sec.VI is a brief summary.  

%%%%%%%%%%%%%%%%%%%%%%%%%%%%%%%%%%%%%%%%%%%%%%%%%%%%%%%%%%%%%%%%%%%%%
\begin{center}
\section{Calogero and Ruijsenaars systems}
\end{center}

%%%%%%%%%%%%%%%%%%%%%%%%%%%%%%%%%%%%%%%%%%%%%%%%%%%%%%%%%%%%%%%%%%%%%
\begin{center}
\subsection{Calogero system}
\end{center}

To begin with, notice that the Seiberg-Witten solution to the SU(2) 
Yang-Mills gauge theory coupled with a massive adjoint matter 
hypermultiplet \cite{DW,IM,DP} is related to the spectral curve 
	\beq
	\det ({\cal{L}}(\xi) -t)=0
	\eeq
of one-dimensional SL(2) elliptic Calogero system with the $2\times2$ Lax 
matrix with Calogero coupling constant $g_0$ 
	\beq
	{\cal{L}}(\xi )=\left(
	\begin{array}{cc}
	P&g_0 \displaystyle\frac{\sigma (Q+\xi)}{\sigma (\xi )\sigma (Q)}\\
	g_0 \displaystyle\frac{\sigma (Q+\xi)}{\sigma (\xi )\sigma (Q)}& -P 
	\end{array}
	\right)
	,\quad P=p_1 =-p_2 ,\quad Q=q_1 -q_2 
	,\eeq
where $p_i$ and $q_i$ are the canonical coordinate and momentum, 
respectively, $\sigma$ is the Weierstrass's $\sigma$ function 
(see Appendix A) and $\xi$ is the spectral parameter associated with an 
elliptic curve 
	\beq
	\quad y^2 =\prod_{i=1}^3 (x-e_i )
	,\quad \sum_{i=1}^3 e_i =0
	.\lab{2.2}
	\eeq
In (\ref{2.2}), branching points $e_i$ are functions in the modulus $\tau$ 
of (\ref{2.2}) and have the expansion
	\beqa
	e_1 &=&\frac{2}{3}(1+24q^2 +24q^4 +\cdots ),\nm\\
	e_2 &=&-\frac{1}{3}(1+24q+24q^2 +96q^3 +\cdots ),\nm\\
	e_3 &=&-\frac{1}{3}(1-24q+24q^2 -96q^3 +\cdots )
	,\eeqa
where $q=e^{i\pi \tau}$. Since $\tau$ is identified with the bare effective 
coupling $\tau=4\pi i/g^2 +\theta/2\pi$, where $g$ is the gauge coupling constant 
and $\theta$ is the vacuum angle, the factor $q^{2n}$ for $n\in \mbox{\boldmath$N$}$ 
corresponds to the instanton amplitude. 

This spectral equation can be summarized into 
	\beq
	 g_{0}^2 \wp (\xi)=t^2 -h ,\quad h=P^2 +\wp (Q)
	\eeq
where $\wp (\xi )$ is the Weierstrass's $\wp$ function and $h$ is the second 
Hamiltonian of this system. 

Then the Seiberg-Witten differential $dS_{\mbox{\scriptsize Cal}}$ is given 
in the form of twice of a product of the eigenvalue $t$ of the Lax matrix 
and a holomorphic 1-form $d\omega =dx/y$ on (\ref{2.2}), namely, 
	\beq
	dS_{\mbox{\scriptsize Cal}}=2 td\omega =
	\frac{2\sqrt{h +g_{0}^2 x}}{y}dx 
	,\eeq
where we have identified $x =\wp (\xi )$. In general, Seiberg-Witten 
differential has the property such that it reduces to a holomorphic 
differential by a differentiation over moduli, and in fact for the case at 
hand, $\pa dS_{\mbox{\scriptsize Cal}} /\pa h \propto dx/y$. Below, 
we set $g_0 =1$ for convenience. 

For this differential, the Seiberg-Witten periods can be defined by 
	\beq
	a=\oint_{\alpha} dS_{\mbox{\scriptsize Cal}},\quad 
	a_D =\frac{\pa \widetilde{\F}}{\pa a}=
	\oint_{\beta} dS_{\mbox{\scriptsize Cal}}
	,\eeq
where $\alpha$ and $\beta$ are the canonical basis of 1-cycles and 
$\widetilde{\F}$ is the prepotential.

Itoyama and Morozov \cite{IM} made a very interesting observation with 
respect to this Seiberg-Witten solution, which states that the data can be 
viewed as if they were given on the hyperelliptic curve 
	\beq
	\widehat{y}^2 =(h +x)\prod_{i=1}^3 (x-e_i )
	\lab{22}
	\eeq
and the associated Seiberg-Witten differential
 	\beq
	dS_{\mbox{\scriptsize Cal}} =\frac{2(h+x)}{\widehat{y}}dx
	\lab{23}
	\eeq
on (\ref{22}). This observation is very important throughout the paper. 

%%%%%%%%%%%%%%%%%%%%%%%%%%%%%%%%%%%%%%%%%%%%%%%%%%%%%%%%%%%%%%%%%%%%%
\begin{center}
\subsection{Ruijsenaars model}
\end{center}

Next, let us discuss the case of one-dimensional SL(2) Ruijsenaars 
model, \cite{IM,Rui} whose Lax operator matrix is given by 
	\beq
	L (\xi )=\sqrt{\frac{\wp (\mu )-\wp (Q)}{\wp (\mu )-\wp (\xi )}}
	\left(\begin{array}{cc}
	e^{P}&\displaystyle 
	e^{P}\frac{\sigma (Q+\xi )\sigma (\mu)}{\sigma (Q+\mu )
	\sigma(\xi )}\\
	\displaystyle e^{-P}\frac{\sigma (-Q+\xi )\sigma (\mu)}
	{\sigma (-Q+\mu )\sigma(\xi )}&e^{-P}
	\end{array}\right)
	.\eeq
The Calogero model is recovered for small $\mu$. \cite{Rui} 
Then the spectral equation
	\beq
	\det (L (\xi )-t )=0
	\eeq
determines the eigenvalue $t$, and by this we can construct it's 
corresponding Seiberg-Witten solution, but in contrast with the preceding 
Calogero model, the Seiberg-Witten differential in this case must take the 
form 
	\beq
	dS_{\mbox{\scriptsize Rui}}
	=\ln t d\omega
	,\quad 
	t=\frac{H\pm \sqrt{H^2 -\wp (\mu )+\wp (\xi )}}{
	\sqrt{\wp (\mu )-\wp (\xi )}},\quad 
	H=\sqrt{\wp (\mu )-\wp (Q)}\cosh P
	\lab{33}
	,\eeq
where $H$ is the Hamiltonian of the system, because of the requirement such 
that $\pa_H dS_{\mbox{\scriptsize Rui}}$ must be a holomorphic 
differential. Below, we take $+$ sign for the eigenvalue $t$ in (\ref{33}). 

The Seiberg-Witten differential of the form (\ref{33}) is 
very characteristic, and takes just the same form with those arising in 
five-dimensional gauge theories. Accordingly, the Ruijsenaars model will be 
understood in the context of higher dimensional gauge theory, although a 
four-dimensional point of view, i.e., 
$dS_{\mbox{\scriptsize Rui}}=2td\omega$, was discussed by Itoyama and 
Morozov. \cite{IM} In fact, as is obvious from (\ref{33}), we can see 
the relation between holomorphic differentials in ``four'' and five dimensions  
	\beq
	\frac{\pa dS_{\mbox{\scriptsize Rui}}}{\pa H}=
	\frac{\pa d\widehat{S}_{\mbox{\scriptsize Cal}}}{\pa 
	\widehat{h}}
	,\lab{35}
	\eeq
where 
	\beq
	d\widehat{S}_{\mbox{\scriptsize Cal}}=
	\frac{2\sqrt{\widehat{h}+x }}{y}dx ,\quad 
	\widehat{h}=H^2 -\wp (\mu )
	.\eeq
Notice that the difference between $d\widehat{S}_{\mbox{\scriptsize Cal}}$ 
and $dS_{\mbox{\scriptsize Cal}}$ is simply $\widehat{h}\leftrightarrow h$. 

For this $dS_{\mbox{\scriptsize Rui}}$, we define the periods 
	\beq
	A=\frac{1}{2i\pi R}\oint_{\alpha}dS_{\mbox{\scriptsize Rui}}=
	\frac{1}{i\pi R}\int_{e_3}^{e_2}dS_{\mbox{\scriptsize Rui}},\quad 
	A_D =\frac{\pa \F}{\pa A}=
	\frac{1}{2i\pi R}\oint_{\beta}dS_{\mbox{\scriptsize Rui}}
	=\frac{1}{i\pi R}\int_{e_2}^{e_1}dS_{\mbox{\scriptsize Rui}}
	,\lab{37}
	\eeq
where we have introduced the prepotential $\F$. Furthermore, 
we have normalized such that these match with those used by Braden 
{\em et al.}, \cite{BMMM} therefore, $R$ plays the role of the radius of 
$S^1$ when this model is regarded as the Seiberg-Witten solution related to the 
five-dimensional gauge theory compactified on $S^1$. 
Then the effective coupling constant is given by 
	\beq
	\tau_{\mbox{\scriptsize eff}}=\frac{\pa A_D}{\pa A}
	.\eeq

%%%%%%%%%%%%%%%%%%%%%%%%%%%%%%%%%%%%%%%%%%%%%%%%%%%%%%%%%%%%%%%%%%%%%
\begin{center}
\section{Pakuliak-Perelomov equations and Picard-Fuchs equations}
\end{center}

To calculate the prepotential of Ruijsenaars model, the periods (\ref{37}) 
should be evaluated, but the periods of a Riemann surface are known to 
satisfy Picard-Fuchs equations. 
In the case at hand, the Picard-Fuchs equation is available from that of the 
Calogero model by focusing on (\ref{35}). As a matter of fact, though the 
Seiberg-Witten Riemann surface (\ref{22}) is hyperelliptic type represented 
by branching points, since the Seiberg-Witten differential (\ref{23}) is a 
linear sum of Abelian differentials on (\ref{22}), we can use a general 
technique to get Picard-Fuchs equations for periods of hyperelliptic Riemann 
surfaces. \cite{IMNS,Ali,IS} 

However, for the case at hand, since the curve (\ref{22}) is given by using 
branching points, the idea of derivation of Picard-Fuchs equations will 
naturally overlap to that of Pakuliak and Perelomov, \cite{PP} provided the 
branching points also including $x=-h$ are regarded as if they were 
independent variables. Namely, Picard-Fuchs equation in the Calogero 
system, and thus that of the Ruijsenaars model, are obtained from 
Pakuliak-Perelomov equations. 

%%%%%%%%%%%%%%%%%%%%%%%%%%%%%%%%%%%%%%%%%%%%%%%%%%%%%%%%%%%%%%%%%%%%%
\begin{center}
\subsection{Pakuliak-Perelomov equations}
\end{center}

In general, the hyperelliptic curve of genus $r$ can be realized as a 
double cover of a polynomial in $x$ 
	\beq
	y^2 =\prod_{i=1}^{2r+2}(x-e_i )=\sum_{i=0}^{2r+2}(-1)^i \sigma_i 
	(e_j ) x^{2r+2-i} 
	\lab{hye}
	,\eeq
where $e_k$ are the branching points on the $x$-plane and we have expressed 
the coefficients of powers in $x$ by $\sigma_i (e_j )$. Do not confuse them 
with the Weierstrass's $\sigma$ function. 

We can define the periods of Abelian differentials on (\ref{hye}) by 
	\beq
	K_j =\oint_{\gamma}\frac{x^{j}}{y}dx
	,\quad j=0,1,\cdots ,2r 
	\lab{s}
	\eeq
where $\gamma$ is an arbitrary non-contractible 1-cycle on (\ref{hye}). 
Pakuliak-Perelomov equations are the equations of a system of 
first-order differential equations satisfied by $K_j$ and the reduction 
method to get such equations can be done in the following way.  

First, notice that the first-order derivatives are given by 
	\beq
	\frac{\pa K_j}{\pa e_i}=\frac{1}{2}\oint_{\gamma}\frac{x^{j}dx}
	{y(x-e_i)}
	.\lab{sss}
	\eeq
Here, defining the integrand as 
	\beq
	I_j =\frac{x^j}{y(x-e_i )}
	,\eeq	
we get the recursion relation
	\beq
	I_j =K_{j-1}+e_i I_{j-1}
	,\lab{rec}
	\eeq
which indicates 
	\beq
	I_j =\sum_{n=0}^{j-1}e_{i}^{j-n-1}K_{n}+e_{i}^{j}I_0
	.\eeq
Thus, from (\ref{sss}), we get 
	\beq
	2\frac{\pa K_j}{\pa e_i}=\sum_{n=0}^{j-1}e_{i}^{j-n-1}K_n +
	e_{i}^j I_0
	.\lab{ssss}
	\eeq

Next, the relation 
	\beq
	\oint_{\gamma}\frac{d}{dx}\left(\frac{y}{x-e_i }\right)dx =0
	\eeq
induces 
	\beq
	\frac{1}{2}\sum_{k=1}^{2r+2}\oint_{\gamma}\frac{P^{(i,k)}(x)}{y}dx=
	\oint_{\gamma}\frac{P^{(i)}(x)}{y(x-e_i )}dx
	,\lab{in}
	\eeq
where 
	\beqa
	& &P^{(i,k)}(x)=\prod_{i,k\neq j=1}^{2r+2}(x-e_j )=
	\sum_{j=0}^{2r}(-1)^j \widehat{\sigma}_{j}^{(i,k)}x^{2r-j},\nm\\ 
	& &P^{(i)}(x)=\prod_{i\neq j=1}^{2r+2}(x-e_j )=
	\sum_{j=0}^{2r+1}(-1)^j \widehat{\sigma}_{j}^{(i)}x^{2r+1-j}
	.\lab{DarthVader}
	\eeqa
In (\ref{DarthVader}), we have expressed the coefficients by 
$\widehat{\sigma}_{j}^{(i)}$ and $\widehat{\sigma}_{j}^{(i,k)}$. 

Again using (\ref{rec}), we can obtain 
	\beq
	\oint_{\gamma}\frac{P^{(i)}(x)}{y(x-e_i )}dx=\sum_{j=0}^{2r+1}(-1)^j 
	\widehat{\sigma}_{j}^{(i)}\left[\sum_{n=0}^{2r-j}e_{i}^{2r-j-n}
	K_{n}+e_{i}^{2r+1-j}\oint_{\gamma}\frac{dx}{y(x-e_i )}\right]
	,\eeq
but from (\ref{ssss}), the Pakuliak-Perelomov equations follow  
	\beqa
	2\frac{\pa K_j}{\pa e_i}&=&\sum_{n=0}^{j-1}
	e_{i}^{j-n-1}K_{n}\nm\\
	& &+\frac{e_{i}^{j}}{P^{(i)}(e_i)}\left[
	\frac{1}{2}\sum_{k=1}^{2r+2}\sum_{n=0}^{2r}(-1)^{2r-n}\widehat{
	\sigma}_{2r-n}^{(i,k)}K_{n}-\sum_{j=0}^{2r+1}\sum_{n=0}^{2r-j}
	(-1)^j \widehat{\sigma}_{j}^{(i)}e_{i}^{2r-j-n}K_{n}\right]
	.\lab{pp}
	\eeqa
The right hand side of (\ref{pp}) is a linear sum of various $K_n$, but 
it would be easy to obtain the equations satisfied by a single $K_i$ by 
repeating differentiations, and accordingly, such equations compose a 
Picard-Fuchs system. 

%%%%%%%%%%%%%%%%%%%%%%%%%%%%%%%%%%%%%%%%%%%%%%%%%%%%%%%%%%%%%%%%%%%%%
\begin{center}
\subsection{Picard-Fuchs equation for Calogero model}
\end{center}

In the Calogero model, if we focus only on $h$-derivatives, the Picard-Fuchs 
equation of the third-order 
	\beq
	4\pa_h (\Del_{\mbox{\scriptsize Cal}} 
	\pa_{h}^2 dS_{\mbox{\scriptsize Cal}})+
	3h\pa_h dS_{\mbox{\scriptsize Cal}} =0
	,\lab{pppf}
	\eeq
where 
	\beq
	\Del_{\mbox{\scriptsize Cal}}(h)=\prod_{i=1}^3 (h+e_i )
	,\eeq
follows from (\ref{pp}). 

Since the Seiberg-Witten solution involves another parameter, 
the bare coupling constant, one more equation including $\tau$-derivatives 
like that discussed by Itoyama and Morozov \cite{IM} may be expected. 
However, the derivation of prepotential from such equation requires 
technical problems, so we do not discuss it in this paper. 

%%%%%%%%%%%%%%%%%%%%%%%%%%%%%%%%%%%%%%%%%%%%%%%%%%%%%%%%%%%%%%%%%%%%%
\begin{center}
\subsection{Picard-Fuchs equation for Ruijsenaars model}
\end{center}

It is now easy to find Picard-Fuchs equation for Ruijsenaars model, if we 
notice the relation (\ref{35}). Since 
$d\widehat{S}_{\mbox{\scriptsize Cal}}$ satisfies the Picard-Fuchs 
equation (\ref{pppf}) replaced $h$ by $\widehat{h}$, we can obtain from 
(\ref{35}) and (\ref{pppf})
	\beq
	\left[\Del \pa_{H}^3 +\Del 
	\pa_H \left(\ln \frac{\Del}{H}\right)\pa_{H}^2 
	+3H^2 \widehat{h}\pa_H \right]dS_{\mbox{\scriptsize Rui}}=0
	,\lab{pfrui}
	\eeq
where 
	\beq
	\Del =\prod_{i=1}^3 \left(\widehat{h}+e_i \right)
	.\eeq

%%%%%%%%%%%%%%%%%%%%%%%%%%%%%%%%%%%%%%%%%%%%%%%%%%%%%%%%%%%%%%%%%%%%%
\begin{center}
\section{Periods}
\end{center}

To derive the instanton correction for the prepotential of Ruijsenaars 
model, some items should be prepared appropriately. The first one is the 
calculation of periods, but in contrast with usual cases, the evaluation of 
periods is very sensitive because the Seiberg-Witten differential of 
Ruijsenaars model involves a normalization factor differential depending on 
$q$, i.e., the second term of the right hand side of 
	\beq
	dS_{\mbox{\scriptsize Rui}}=\ln\left[ H+\sqrt{\widehat{h}+x}
	\,\right]\frac{dx}{y}-\ln\Bigl[ \wp (\mu )+x\Bigr]\frac{dx}{y}
	.\eeq
Since this term is independent of $H$, the Picard-Fuchs equation can not 
detect the contribution of instantons arising from this term. In fact, 
this is because $\wp (\mu )$ can be expanded by $q$. Therefore, it is 
necessary to calculate this in order to include instanton effects correctly. 
Otherwise, the prepotential will not reduce to any physical prepotential by 
scaling limits. \cite{BMMM} 

With this in remind, we can see that the period $A$ in the weak coupling 
region ($i\tau\rightarrow \infty$, $q=e^{i\pi \tau} \rightarrow 0$) behaves 
as
	\beq
	iRA=\ln \left[ H\sin \mu +\sqrt{H^2\sin^2 \mu -1}\,\right]-\frac{4
	H(H^2 \sin^2 \mu -\sin^2 \mu -1)\sin^3 \mu }{(H^2 \sin^2 \mu -1)^{
	3/2}}q^2 +\cdots 
	\eeq
and thus it's inverse relation follows immediately 
	\beq
	H=\frac{\cos RA}{\sin \mu}+\frac{2(2-\cos 2\mu -\cos 2RA)
	\sin \mu \cos RA}{\sin^2 RA}q^2 
	+\cdots  
	.\lab{riv}
	\eeq

On the other hand, as for the dual period $A_D$, it is enough to calculate 
it only at the perturbative level for a later convenience 
(see also Ref.24). 

	\beqa
	\frac{\pa A_{D}}{\pa H}\Biggr|_{q\rightarrow 0}&=&-\frac{1}{\pi R}
	\lim_{\epsilon \to 0}\int_{\epsilon}^{\infty}
	\frac{dx}{x\sqrt{x(\widehat{h}+2/3)+\widehat{h}-1/3}}\nm\\
	&=&-\frac{1}{\pi R}\frac{\sin\mu}{\sqrt{
	H^2 \sin^2 \mu -1}}\left[\ln\frac{H^2 \sin^2 \mu -1}{
	H^2 \sin^2 \mu -\cos^2 \mu}-\ln\frac{\epsilon}{4}\right]\Biggr|_{
	\epsilon \rightarrow 0}
	.\lab{ep}
	\eeqa

Accordingly, extracting the finite part of (\ref{ep}) and with the help of 
(\ref{riv}), we get the perturbative effective coupling constant 
	\beq
	\tau_{\mbox{\scriptsize eff}}\Bigr|_{q\rightarrow 0}
	=\frac{1}{i\pi}\ln\frac{\sin^2 RA}{\sin^2 \mu
	-\sin^2 RA}
	.\lab{tau}
	\eeq
This coincides with the perturbative calculus. \cite{Nek}

%%%%%%%%%%%%%%%%%%%%%%%%%%%%%%%%%%%%%%%%%%%%%%%%%%%%%%%%%%%%%%%%%%%%%
\begin{center}
\section{Instanton correction for prepotential of Ruijsenaars model}
\end{center}

%%%%%%%%%%%%%%%%%%%%%%%%%%%%%%%%%%%%%%%%%%%%%%%%%%%%%%%%%%%%%%%%%%%%%
\begin{center}
\subsection{Differential equation for prepotential}
\end{center}

In the case of SU(2) gauge group, we can give a 
differential equation for prepotential by using 
a familiar method using inversion relations of periods. \cite{Mat,KO} 

For our case, derivatives of periods are inverted by \cite{KO} 
	\[
	\pa_H A=\frac{1}{H^{\,'}},\quad \pa_{H}^2 A=-\frac{H^{\,''}}
	{H^{\,'3}}
	,\quad \pa_{H}^3 A=3\frac{H^{\,''2}}{H^{\,'5}}-\frac{H^{\,'''}}{
	H^{\,'4}},\]
	\vspace{-1.0cm}
	\beq
	\pa_{H}A_{D} =\frac{\F^{\,''}}{H^{\,'}},\quad 
	\pa_{H}^2 A_{D} =\frac{\F^{\,'''}}{H^{\,'2}}-\frac{\F^{\,''}
	H^{\,''}}{H^{\,'3}},\quad 
	\pa_{H}^3 A_D =\frac{\F^{(4)}}{H^{\,'3}}-3\frac{H^{\,''}\F^{\,'''}}
	{H^{\,'4}}+\left(3\frac{H^{\,''2}}{H^{\,'5}}-\frac{H^{\,'''}}{
	H^{\,'4}}\right)\F^{\,''}
	,\lab{51}
	\eeq
where $' =\pa /\pa A$. 

Then from (\ref{pfrui}), we have 
	\beq
	\F^{(4)}-\left(3\frac{H^{\,''}}{H^{\,'}}-\frac{\pa_H \Del}
	{\Del}\right)\F^{\,'''}=0
	,\eeq	
which is integrated to give 
	\beq
	\F^{\,'''}=c\frac{H H^{\,'3}}{\Del}
	\lab{scale}
	.\eeq
In (\ref{scale}), $c$ is an integration constant to be fixed below. 
Versions of (\ref{scale}) can be found in several SU(2) gauge 
theories. \cite{KO,O5} 
Note that (\ref{scale}) agrees with the equation obtained from 
residue calculus [cf. (44) in Ref.24] in the perturbative limit.

%%%%%%%%%%%%%%%%%%%%%%%%%%%%%%%%%%%%%%%%%%%%%%%%%%%%%%%%%%%%%%%%%%%%%
\begin{center}
\subsection{Instanton correction of prepotential}
\end{center}

Next, substituting (\ref{riv}) into (\ref{scale}) and expanding it by small 
$q$, we can obtain the third-order derivative of the prepotential
	\beq
	\frac{\F^{\,'''}}{cR^3}=
	-\frac{2\cot RA \sin^2 \mu}{\cos 2RA -\cos 2\mu}+
	\frac{F_2 \cot RA}{(\cos 2RA-\cos 2\mu )^2}
	\left(\frac{\sin\mu}{\sin RA}\right)^4 q^2 +\cdots
	,\lab{kkk}
	\eeq
where
	\beqa
	F_2 &=&-36-107\cos 2RA -14\cos 4RA +\cos 6RA -38\cos 4\mu
	+100\cos 2\mu \nm\\
	& &+88\cos 2RA \cos 2\mu +20\cos 4RA \cos 2\mu -14\cos 2RA \cos 4\mu
	.\eeqa
The second term of the right hand side in (\ref{kkk}) is the one-instanton 
contribution. Note that the prepotential is expanded by the invariant 
quantity $q$ under the transformation $\tau \rightarrow \tau +1$. 

To fix the constant $c$, look at the perturbative part of (\ref{kkk}). 
Furthermore, recalling $\F^{\,'''}=\pa \tau_{\mbox{\scriptsize eff}}/\pa A$, 
we get 
	\beq
	c=i\frac{2}{\pi R^2}
	\eeq	
from (\ref{tau}). In this way, we can arrive at the exact expression of 
the third-order derivative of prepotential.  

%%%%%%%%%%%%%%%%%%%%%%%%%%%%%%%%%%%%%%%%%%%%%%%%%%%%%%%%%%%%%%%%%%%%%
\begin{center}
\subsection{Reduction to Calogero prepotential}
\end{center}

As a check of our calculus, let us take a scaling limit 
$\mu =2i\pi g_0 R, R\rightarrow 0$ for $\F^{\,'''}$. 
Then it is straightforward to see that  
	\beq
	\F^{\,'''}=-\frac{8i\pi g_{0}^2}{A(A^2 +4\pi^2 g_{0}^2 )}
	-\frac{256i\pi^3 g_{0}^4 (-135A^4 +2520\pi^2 A^2 g_{0}^2 +9360\pi^4 
	g_{0}^4 )}{45A^5 (A^2 +4\pi^2 g_{0}^2 )^2}q^2 +\cdots
	.\lab{ten}
	\eeq
The first term of the right hand side of (\ref{ten}) can be identified with 
the perturbative part of the Calogero model, but the second one does not 
seem to be that of the one-instanton contribution of the Calogero model. 
However, expanding it for a small $g_0$, we get 
	\beq
	\F^{\,'''}=-\frac{8i\pi g_{0}^2}{A(A^2 +4\pi^2 g_{0}^2 )}
	+\left( \frac{768i\pi^3}{A^5}g_{0}^4 -\frac{20480i\pi^5}{A^7}g_{0}^6 
	+\cdots \right)q^2 +\cdots
	,\lab{tten}
	\eeq
whose first term in the brackets is nothing but the one-instanton 
correction of the Calogero model (see Appendix B), 
provided higher order terms in $g_0$ are ignored (this operation is 
necessary because the original reduction to the Calogero model is 
supplied by $\mu \rightarrow 0$, and then this is equivalent to the 
assumption of a small mass of the adjoint hypermultiplet). 
Actually, though the sign is different, 
this is due to the ambiguity of the Weyl reflection for $a$. Therefore, 
we can conclude that (\ref{tten}) coincides with the third-order derivative 
of the prepotential of the Calogero model. 
Other scaling limits \cite{BMMM} can be treated in a similar way. 

%%%%%%%%%%%%%%%%%%%%%%%%%%%%%%%%%%%%%%%%%%%%%%%%%%%%%%%%%%%%%%%%%%%%%
\begin{center}
\section{Summary}
\end{center}

In this paper, we have shown that:
	\begin{itemize}
	\item
	The Picard-Fuchs equation in the Ruijsenaars system is generated 
	from the Pakuliak-Perelomov equations. 
	
	\item
	From the differential equation for prepotential, one-instanton 
	prepotential of the Ruijsenaars model is obtained. 
 
	\item	
	Our prepotential is checked in the limit to the Calogero system, 
	i.e., the $N=2$ gauge theory with a massive adjoint hypermultiplet. 
	
	\end{itemize}

%%%%%%%%%%%%%%%%%%%%%%%%%%%%%%%%%%%%%%%%%%%%%%%%%%%%%%%%%%%%%%%%%%%%%
\begin{center}
\section*{Appendix A: Weierstrass functions}
\end{center}

\renewcommand{\theequation}{A\arabic{equation}}\setcounter{equation}{0}

Suppose that $f(z)$ is a complex single valued function with the 
complex argument $z$. Then if $f(z+\omega )=f(z)$ holds for a complex number 
$\omega$, $f$ is called a periodic function with period $\omega$. If $f$ 
has two independent periods $\omega$ and $\widehat{\omega}$, $f$ is 
referred to double periodic. In addition, if $f$ is rational, 
$f$ is called elliptic function. The Weierstrass functions are the 
elliptic functions with the following properties. 
	\begin{itemize}
	\item
	Definition:
	\beq
	\wp (z)=\frac{1}{z^2}+\sum_{-\infty <m=n\neq 0<\infty}\left[
	\frac{1}{(z+m\omega +n\widehat{\omega})^2}-\frac{1}{
	(m\omega +n\widehat{\omega})^2}\right]
	.\eeq
	
	\item
	Relation between $\wp$ and $\sigma$: 
	\beq
	\wp (z)=\frac{\pa_z \sigma (z)}{\sigma (z)},
	\quad \wp (z)-\wp (\xi )=-\frac{\sigma(z+\xi)\sigma(z-\xi)}
	{[\sigma(z)\sigma(\xi)]^2}
	.\eeq

	\item
	Parity: 
	\beq
	\wp(z)=\wp (-z),\quad \sigma (z)=-\sigma (z).
	\eeq
	
	\item
	Differential equation:
	\beq
	\left[\pa_z \wp (z)\right]^2 =4\prod_{i=1}^{3}\Bigl[\wp (z)-e_i \Bigr]
	.\eeq

	\item
	Laurent expansion:
	\beq
	\wp (z)=\frac{1}{z^2}+\frac{g_2}{20}z^2 +\frac{g_3}{28}z^4 +
	\cdots,\quad -\frac{g_2}{4}=e_1 e_2 +e_2 e_3 +e_3 e_1 ,\quad 
	\frac{g_3}{4}=e_1 e_2 e_3  
	.\eeq
	
	\item
	$q=e^{i\pi \tau}$-expansion: 
	\beq
	\wp (z)=-\frac{1}{3}+\frac{1}{\sin^2 z}+16q^2 \sin^2 z +\cdots
	.\eeq 
	\end{itemize}

%%%%%%%%%%%%%%%%%%%%%%%%%%%%%%%%%%%%%%%%%%%%%%%%%%%%%%%%%%%%%%%%%%%%%
\begin{center}
\section*{Appendix B: Prepotential of Calogero model}
\end{center}

\renewcommand{\theequation}{B\arabic{equation}}\setcounter{equation}{0}

In this appendix, we summarize some formulae and prepotential in Calogero 
model, but the derivation of prepotential is parallel to that of 
Ruijsenaars model, so only the necessary items are listed, and $g_0$ is 
identified as $g_{0}^2 =im^2 /\pi$, where $m$ is the mass of the adjoint 
hypermultiplet. In this calculation, we present in the weak coupling 
region, and the 1-cycles are taken as the same ones with the Ruijsenaars 
model. In addition, note that due to our normalization there are several 
distinctions to the known results. \cite{IM,DP}  
	\begin{itemize}
	\item
	Periods and inverse relation ($\widetilde{h}=h/g_{0}^2$): 
	\beqa
	\frac{a}{g_0 \pi}&=&\frac{2}{3}\sqrt{9\widetilde{h}-3}+\frac{
	8\sqrt{3}(3\widetilde{h}-4)}{(3\widetilde{h}-1)^{3/2}}q^2 +
	\cdots,\nm\\
	\frac{\pa a_D}{\pa \widetilde{h}}&=&-
	\frac{ig_0}{\sqrt{\widetilde{h}-1/3}}\left[2\ln 2+
	\ln \frac{3\widetilde{h}-1}{3\widetilde{h}+2}\right]+\cdots,\nm\\
	\widetilde{h}&=&\frac{1}{3}+\frac{a^2}{4\pi^2 g_{0}^2}+\left(-8+
	\frac{32\pi^2 g_{0}^2}{a^2}\right)q^2 +\cdots
	.\eeqa

	\item
	Effective coupling constant:
	\beq
	\frac{\pa^2 \widetilde{\F}}{\pa a^2}\Biggr|_{q\rightarrow 0}=
	\frac{i}{\pi}\ln\left(\frac{1}{4}+\frac{\pi^2 g_{0}^2}{a^2}\right)
	.\eeq

	\item
	Third-order derivative of prepotential: 
	\beqa
	\frac{\pa^3 \widetilde{\F}}{\pa a^3}&=&-i\pi g_{0}^2
	\frac{(\pa_a \widetilde{h})^3}{\Del_{\mbox{\scriptsize Cal}}
	(\widetilde{h})}\nm\\
	&=&-\frac{8i\pi g_{0}^2}{a(a^2 +4\pi^2 g_{0}^2 )}-
	\frac{768i\pi^3 g_{0}^4}{a^5}q^2 +\cdots
	.\eeqa

	\end{itemize}

%%%%%%%%%%%%%%%%%%%%%%%%%%%%%%%%%%%%%%%%%%%%%%%%%%%%%%%%%%%%%%%%%%%%
\begin{center}

\end{center}

\end{document}